\documentclass[conference]{IEEEtran}
\IEEEoverridecommandlockouts
\usepackage{cite}
\usepackage{amsmath,amssymb,amsfonts}
\usepackage{algorithmic}
\usepackage{graphicx}
\usepackage{textcomp}
\usepackage{multirow}
\usepackage{listings}
\usepackage{caption}
\usepackage{subcaption}

\usepackage{xcolor}
\def\BibTeX{{\rm B\kern-.05em{\sc i\kern-.025em b}\kern-.08em
    T\kern-.1667em\lower.7ex\hbox{E}\kern-.125emX}}
\begin{document}

\definecolor{codegreen}{rgb}{0,0.6,0}
\definecolor{codegray}{rgb}{0.5,0.5,0.5}
\definecolor{codepurple}{rgb}{0.58,0,0.82}
\definecolor{backcolour}{rgb}{0.95,0.95,0.92}

\lstdefinestyle{mystyle}{
    backgroundcolor=\color{backcolour},   
    commentstyle=\color{codegreen},
    keywordstyle=\color{magenta},
    numberstyle=\tiny\color{codegray},
    stringstyle=\color{codepurple},
    basicstyle=\ttfamily\footnotesize,
    breakatwhitespace=false,         
    breaklines=true,                 
    captionpos=b,                    
    keepspaces=true,                 
    numbers=left,                    
    numbersep=5pt,                  
    showspaces=false,                
    showstringspaces=false,
    showtabs=false,                  
    tabsize=2
}

\lstset{style=mystyle}


\title{Simulation-Based Performance Prediction of HPC Applications: A Case Study of HPL}

\author{\IEEEauthorblockN{Gen Xu\IEEEauthorrefmark{1},
Huda Ibeid\IEEEauthorrefmark{4}, Xin Jiang\IEEEauthorrefmark{1},
Vjekoslav Svilan\IEEEauthorrefmark{4}, and Zhaojuan Bian\IEEEauthorrefmark{1}}
\IEEEauthorblockA{\IEEEauthorrefmark{1} Intel Corporation, Shanghai, China}
\IEEEauthorblockA{\IEEEauthorrefmark{4} Intel Corporation, Santa Clara, US}
\{gen.xu, huda.ibeid, xin.jiang, vjekoslav.svilan, bianny.bian\}@intel.com
}

\maketitle

\begin{abstract}
We propose a simulation-based approach for performance modeling of parallel applications on high-performance computing platforms. Our approach enables full-system performance modeling: (1) the hardware platform is represented by an abstract yet high-fidelity model; (2) the computation and communication components are simulated at a functional level, where the simulator allows the use of the components native interface; this results in a (3) fast and accurate simulation of full HPC applications with minimal modifications to the application source code. This hardware/software hybrid modeling methodology allows for low overhead, fast, and accurate exascale simulation and can be easily carried out on a standard client platform (desktop or laptop). We demonstrate the capability and scalability of our approach with High Performance LINPACK (HPL), the benchmark used to rank supercomputers in the TOP500 list. Our results show that our modeling approach can accurately and efficiently predict the performance of HPL at the scale of the TOP500 list supercomputers. For instance, the simulation of HPL on Frontera takes less than five hours with an error rate of four percent.
\end{abstract}


\section{Introduction}
Currently, there are many efforts to evaluate the hardware and software bottlenecks of exascale designs to enable the development of applications that exploit the full performance of exascale computing platforms. However, the increasing complexity of modern computing architectures along with the exponentially growing configuration space and complex interactions among configuration options often make it difficult to develop accurate performance models. In recent years there have been several efforts to model the performance of HPC applications using simulation-based approaches. However, several challenges must be addressed to enable these approaches.

The full system stack consists of three layers: hardware infrastructure, middle layer libraries, and the application itself. Each layer can have a huge impact on the overall performance, which means that all layers should be modeled to acheive an acceptable accuracy. One of the main challenges is to determine which aspects are the most important to simulate when modeling each layer for large scale HPC applications. In terms of the hardware infrastructure layer, computation components, such as CPU, GPU, and memory, should be modeled. Similarly, the interconnect network is one of the essential parts. The computation and communication platforms are the most important to take into consideration for the distributed system.

Choosing which libraries to simulate is another important aspect. The basic principle is to choose the most widely used libraries. Science and engineering computations have been the dominant category of the applications running on HPC systems. In this area, Basic Linear Algebra Subprograms~\cite{blackford_updated_2002} (BLAS) is the most widely used mathematical library that forms the computational core of many HPC applications. BLAS operations very time-consuming as well as compute-intensive. Additionally, Message Passing Interface (MPI) has now emerged as the de-facto standard for node-to-node communication on supercomputers. MPI standards are used on all leading supercomputers of the TOP500 list~\cite{gropp1999mpi}. Taking the charactaristics of the software libraries is an essential requirement for accurate simulation-based modeling.

With the hardware infrastructure and software libraries models, our goal is to enable the modeling of HPC applications with minimal modification to the application source code. Among all HPC applications, the High-Performance LINPACK (HPL) Benchmark is the most widely recognized metric for ranking HPC systems, although other benchmarks such as HPGMG~\cite{adams2014} and HPCG~\cite{dongarra2015} have been proposed as either alternative or complementary benchmarks.

In this paper, we propose a simulation framework that employs a layered architecture to simulate HPC systems on standard client computers (desktop or laptop). We use HPL to demonstrate the capability and scalability of the simulation framework. The key contributions of this paper are as follows:
\begin{itemize}
\item We present a hardware platform model that includes the processing nodes and the interconnection network. The model employs a stream-level network model that balances the simulation speed and accuracy.
\item We present abstracted library models for BLAS computations and MPI communications.
\item We model HPL benchmark to demonstrate the capability and scalability of our simulation framework.
\item We demonstrate that our modeling approach can accurately and efficiently predict the performance of HPL at the scale of the TOP500 list supercomputers.
\end{itemize}

The rest of the paper is organized as follows. In section~\ref{sec:related}, we present a background on simulation-based approaches. We also describe related work in hardware infrastructure simulation and MPI modeling. In section~\ref{sec:arch}, we present an overview of our simulation framework and describe the design of each of its layers. In section~\ref{sec:results}, we conduct extensive validation and performance studies. In section~\ref{sec:cases} we present some use cases. Finally, conclusions and future directions are presented in section~\ref{sec:conclusion}.

\section{Background and Related work}
\label{sec:related}

In recent years there have been several efforts to model the performance of HPC applications using simulation-based approaches.

SimGrid~\cite{casanova_simgrid_2008} is an open-source simulation framework for large-scale distributed systems. It was originally designed to study the behavior of Grids but has been extended and applied to a wide range of distributed computing platforms, including Clouds and High Performance Computing systems. SimGrid uses a flow-level approach that approximates the behavior of TCP networks. Due to its use of a flow-level network simulation approach along with a coarse-grained CPU model for the computation, SimGrid can perform large numbers of statistically significant experiments on large TCP networks. However, SimGrid might result in an unacceptable accuracy when compared to packet-level simulators when the data sizes are small or when networks are highly contended~\cite{fujiwara2007speed}. In addition, the lack of detailed models for the processing components makes SimGrid unsuitable for several HPC applications.

The Structural Simulation Toolkit (SST)~\cite{rodrigues_structural_2011} enables the co-design of highly concurrent systems by allowing simulation of diverse aspects of the hardware and software. SST aims to simulate full-scale machines using a coarse-grained simulation approach for the processing and network components through the use of skeleton applications that replicate the full application control flow.

The work presented in this paper builds on our previous work, CSMethod~\cite{bian_simulating_2014}. CSMethod enables full-system performance modeling and prediction of big data clusters by simulating both the software stack (e.g. HDFS, OS, and JVM) and the hardware components (CPU, storage, and network). With CSMethod, the computation and communication behaviors of the application are abstracted and simulated at a functional level. Software functions are then dynamically mapped onto hardware components. To achieve fast and accurate performance simulation, CSMethod supports fine-grained analytical models for processor, memory, and storage. The timing of the hardware components is modeled according to payload and activities as perceived by the software. CSMethod capabilities and accuracy have been demonstrated in~\cite{wang_simulating_2014, wang_millipedes_2015, liu_planning_2016, chen_simulating_2016}. However, CSMethod is focused on big data applications and has not been applied to simulate HPC systems.

Cycle-accurate simulators are commonly used to evaluate next generation processors and system architectures. Traditionally, these simulators trade speed for accuracy. Similarly, packet-level or flit-level network simulators aim for a highly accurate representation of actual network behavior. Thus, large-scale simulations may be too time-consuming with packet-level simulation.

There are several different approaches to model MPI, ranging from analytical models to trace-based simulations. Some MPI modeling frameworks rely on the use of test environments based on ``artificial communications'' to perform synthetic tests of MPI performance. For example, LogGOPSim~\cite{hoefler_loggopsim_2010} replaces MPI collective operations by a set of point-to-point algorithms. While this approach is accurate on smaller systems, LogGOPSim ignores congestion in the network and assumes full effective bisection bandwidth, which may decrease the accuracy of the simulations on emerging large-scale systems. SMPI~\cite{degomme_simulating_2017} simulates unmodified MPI applications on top of the SimGrid simulator. SMPI supports different performance modes through a generalization of the LogGPS model.

\section{Simulation framework}
\label{sec:arch}

\begin{figure}
  \centering
  \includegraphics[width=.8\linewidth]{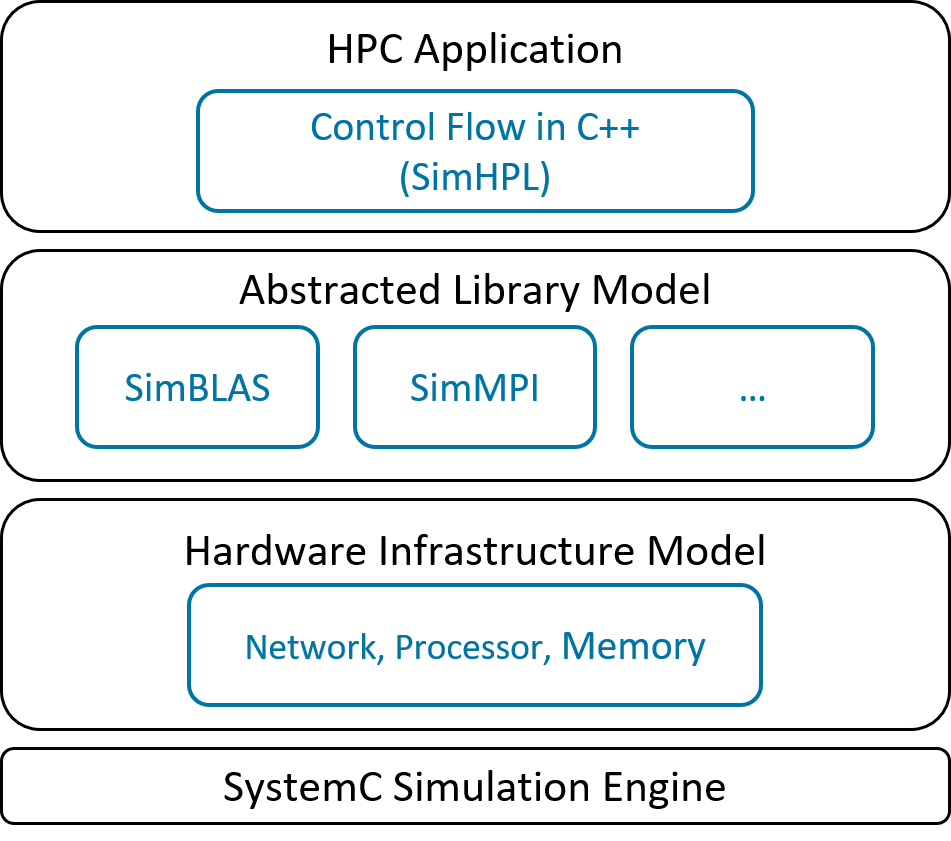}
  \caption{Simulation framework architecture.}
  \label{fig:architecture}
\end{figure}

Our simulation framework employs a layered and configurable architecture to simulate the full stack of supercomputing systems, as shown in Figure~\ref{fig:architecture}. The top layer is the HPC application, where the application behavior is modeled. Underneath the top layer, computation and communication libraries are abstracted and simulated at a functional level. The library layer receives function calls from the top later and dynamically connects to the hardware components. The hardware infrastructure layer beneath the library layer aims at defining the hardware components (processor, network, and storage) of the HPC system. In this framework, software behavior and hardware infrastructure are loosely coupled, which provides the flexibility to change the hardware platform without the need to modify the software behavior model and vice versa.

This paper discusses several techniques: (1) the hardware platform is modeled by an abstract yet high-fidelity model; (2) computation and communication components are simulated at a functional level, where the simulator allows the use of the component native interface; this results in a (3) fast and accurate simulation of HPC applications with minimal modifications to the application source code; and, at the bottom of these layers, (4) a simulation engine for SystemC-based discrete events. This is a low-overhead engine that enables fast simulations with good scalability. This hardware/software hybrid modeling methodology allows for low overhead, fast, and accurate Exascale systems simulation and can be easily carried out on a standard client platform.

\subsection{HPC hardware infrastructure simulation}
\label{sec:hardware_model}
The hardware model builds on our previous work, CSMethod~\cite{bian_simulating_2014}. Here, we extend CSMethod to enable the modeling of HPC applications. In particular, we implement an efficient CPU model for the computation operations as well as a GPU model. Moreover, a stream-level network model is implemented as an alternative to the original packet-level network model.

The hardware model simulates all the main components of the HPC platform, which includes the processing nodes and the interconnection network. In particular, the hardware infrastructure layer consists of models for the CPU, GPU, memory, and NIC. This section describes these models.

\subsubsection{\textbf{Node architecture: CPU, GPU, and memory}}
In this work, we extend~\cite{bian_simulating_2014} to support heterogeneous architectures. This new feature enables the simulation of accelerator-based architectures, such as CPU–GPGPU combinations. Our framework also utilizes analytical models to model compute-bound and bandwidth-bound operations, such as BLAS DGEMM operation and DSWAP described in section~\ref{sec:blas}. Traditionally, compute-bound operations are modeled using an actual single-core execution time on real hardware scaled to the simulated processor core speed. In this work, we model the computation time of these operations analytically based on the theoretical peak performance and the efficiency of these operations on the CPU and GPU. The efficiency can be directly measured without complex computations. Similarly, modeling bandwidth-bound operations is based on the peak bandwidth and bandwidth efficiency.

\subsubsection{\textbf{Interconnection network}}
As discussed earlier, packet-level network models are not suitable for all scenarios. In this work, a stream-level network model is implemented as an alternative that offers latency and bandwidth restrictions. This work extends the capabilities of~\cite{bian_simulating_2014} network model in two ways. First, we include more network architectures, such as fat-tree and dragonfly, which are the most widely used networks in HPC systems. Second, traditionally, the implementation of routing policies calculates and stores all the routing paths during the initialization phase which uses a large amount of memory when simulating large-scale systems. Several routing algorithms, such D-mod-K for fat-tree~\cite{zahavi_d-mod-k_2010} and minimal/non-minimal routing for dragonfly topology~\cite{kim_technology-driven_2008} can be dynamically calculated which reduces the memory consumption significantly.

To model the network communication, we divide large messages into smaller chunks and calculate the transmission time according to the currently allocated bandwidth. In addition, the network model supports communication primitives, such as send data and receive data, which enables the integration of external network simulators into our framework.

\subsection{Computation and communication libraries simulation}
When developing simulation models for large scale complex systems, it is important to consider which components to model. In HPC applications, computation and communication libraries are commonly utilized and tuned for optimal performance. In this work, BLAS and MPI  libraries are simulated as modules on top of the infrastructure layer by leveraging dedicated APIs to access the hardware resources. These modules allow the use of the libraries native interface, thus easing the development of the simulation APIs.

In this section, a detailed discussion of the computation and communication libraries is presented.
 
\subsubsection{\textbf{Performance modeling of BLAS library}}
\label{sec:blas}

Many HPC applications rely heavily on BLAS kernels. The BLAS library implements fundamental dense vector and matrix operations, such as various types of multiplications and triangular linear system solvers. Since these kinds of kernels do not influence the control flow, the simulation time can be reduced by substituting the BLAS function calls with an analytical performance model for the respective kernel. The BLAS operation is data-independent, i.e., the data content does not affect the computation time. This means that all multiplications with zeros are explicitly performed no matter how sparse an operand is (i.e., how few non-zero entries it has).

BLAS functionality is categorized into three sets of levels according to the arithmetic density. Level 1 BLAS operations typically take linear time, $\mathcal{O}(N)$, Level 2 operations quadratic time, and Level 3 operations cubic time. Thus, we employ the same modeling approach but with different analytical performance models that are based on the Roofline model~\cite{williams_Roofline_2009}. The Roofline model provides a simple way to estimate the performance based on the computation kernel and hardware characteristics. It relies on the concept of Arithmetic Intensity (in FLOPs/byte) and provides performance bounds for compute-bound and memory bandwidth-bound computations. 

\textbf{Modeling Level-3 BLAS Kernels:} Here we describe in detail the methodology used to model the DGEMM operation. A similar approachis used to model the DTRSM kernel.

GEMM performs a matrix-matrix multiplication and an add operation
\begin{equation}
C \leftarrow \alpha AB + \beta C,
\end{equation}
where $C$ is $m \times n$, $A$ is $m \times k$, $B$ is $k \times n$, and $\alpha$ and $\beta$ are scalars.

For dense matrices, the total number of operations performed by GEMM is
\begin{equation}
O_{GEMM} = 2mnk+2mn.
\end{equation}

As the GEMM kernel is compute-bound, we use the following analytical model to estimate its compute time
\begin{equation} \label{eq:gemm}
E = \mu O_{GEMM} + \theta,
\end{equation}
where $\mu$ represents the computation cost of a single operation and $\theta$ represents the overhead of each DGEMM function call. The Roofline model sets an upper bound on performance of a kernel depending on its arithmetic intensity. For a more realistic estimates, we take into account the kernel efficiency on a given hardware. Let $e$ be the GEMM efficiency on a given hardware, then $\mu$ can be calculated as the inverse of the multiplication of $e$ by the theoretical peak performance.

\begin{figure}
  \centering
  \includegraphics[width=\linewidth]{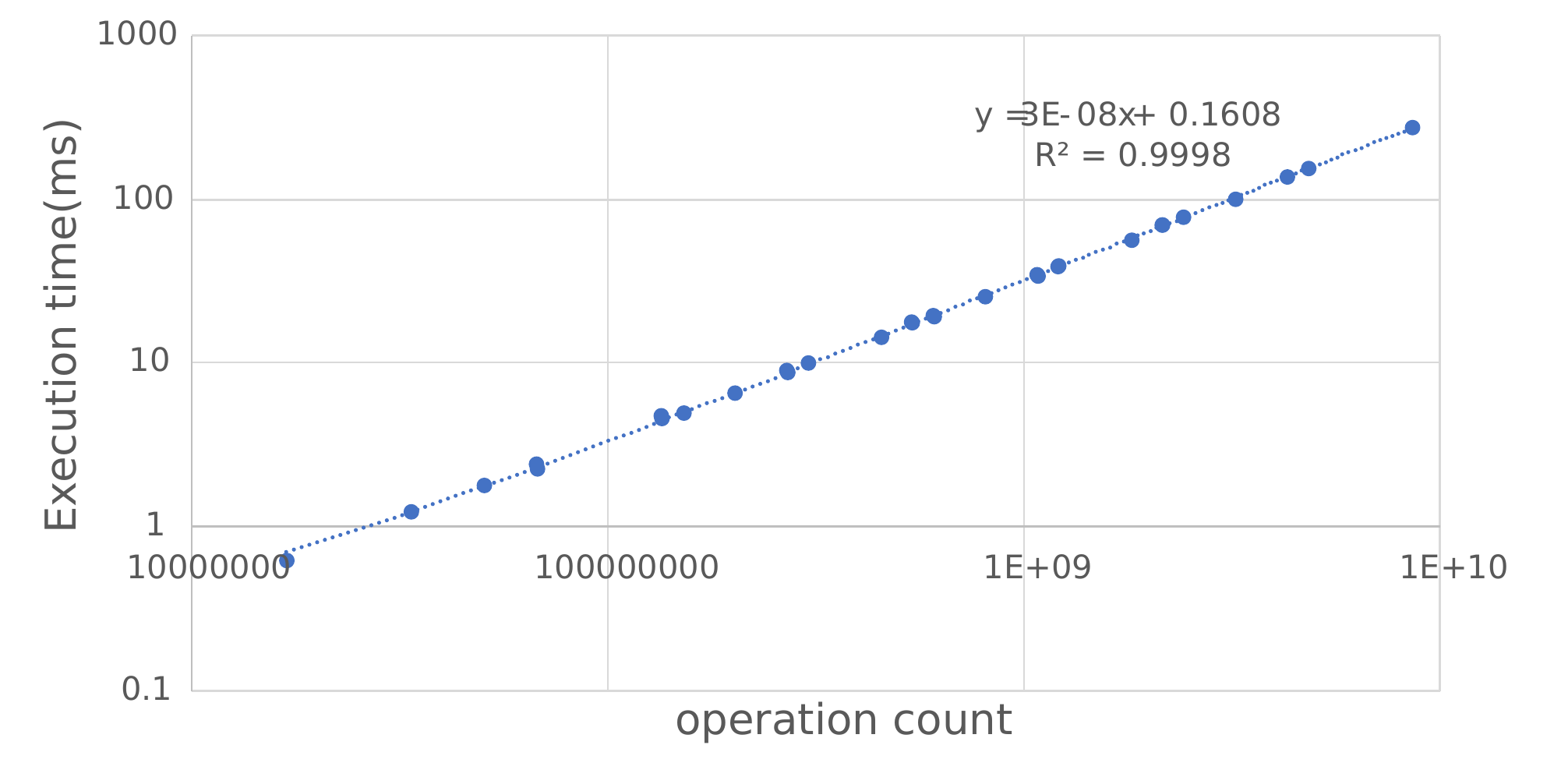}
  \caption{Execution time of DGEMM kernel.}
  \label{fig:DGEMM}
\end{figure}

Both $\mu$ and $\theta$ in~\eqref{eq:gemm} are obtained through profiling and calibration. To calibrate and validate our model, we conduct a micro-test using MKL DGEMM kernel on a single core. The values of $m$, $n$, and $k$ range from $128$ to $2048$. Each case is executed $1000$ times and then the average time is calculated. Figure~\ref{fig:DGEMM} shows the impact of the total number of operations on the execution time along with the estimation model. The validation results show that the R-squared value is $0.9998$. Here, the values of $\mu$ and $\theta$ are implementation and hardware dependent. This kind of analytical modeling speeds up the simulation by orders of magnitude, especially as the matrix size grows.  

\textbf{Modeling Level-1 and Level-2 BLAS Kernels:} A similar approach is employed to model Level-1 and Level-2 BLAS kernels. On most architectures, Level-1 BLAS vector-vector operations, and Level-2 BLAS matrix-vector operations are memory-bound. As mentioned previously, we calibrate the models to take into account the memory efficiency of these operations.

Based on the methodology discussed, we present SimBLAS, a library to simulate and predict the performance of BLAS operations. Figure~\ref{fig:BLAS} shows a code snippet of Level-3 and Level-1 SimBLAS operations. There are different implementations of the BLAS library, for example, cuBLAS for GPUs, OpenBLAS, and Intel BLAS. Each implementation has different efficiency. Furthermore, these implementations can run on a single thread or with multi-threading. Hence, predicting efficiency analytically is a complicated task. In our simulations, we employ a microbenchmark to profile the efficiency and then use it as an input to SimBLAS.
\begin{figure}
  \centering
\begin{lstlisting}[language=c++]
void simblas_dgemm(const SIMBLAS_ORDER Layout, 
                   const SIMBLAS_TRANSPOSE TransA,
                   const SIMBLAS_TRANSPOSE TransB, 
                   const int M, const int N, 
                   const int K, const double alpha, 
                   const double *A, const int lda, 
                   const double *B, const int ldb,
                   const double beta, 
                   double *C, const int ldc) {
    // Number of FLOPS
    double op_count = M * N * (2 * K + 2);
    // Achieved performance
    double perf = getDgemmPerf();
    waitns(op_count/getDgemmPerf() + getBlasLat());    
    return;
}

void simblas_dswap(const int N, 
                   double *X, const int incX, 
                   double *Y, const int incY) {
    // Message size
    double data_movement = 4.0 * N;
    // Achieved performance
    double perf = getDswapPerf();
    waitns(data_movement/perf + getBlasLat());
    return;
}
\end{lstlisting}
  \caption{Implementation of SimBLAS operations.}
  \label{fig:BLAS}
\end{figure}

SimBLAS library is coupled with the underlying hardware models, specifically, CPU, GPU, and memory models. As discussed earlier in this section, the execution time is determined by the operation complexity and hardware characteristics. The operation complexity is the operations count of a compute-bound operation or the memory access size of a bandwidth-bound operation. The hardware characteristics are obtained from the underlying hardware models.

In summary, these performance models, in principle, balance simulation speed and accuracy to predict the performance of HPC systems.

\subsubsection{\textbf{Performance modeling of MPI library}} In our previous work, a set of socket-like APIs are implemented to support TCP network transmission in big data environments. On HPC platforms, MPI is the de-facto standard for inter-node communication. This section details the MPI model in two aspects: peer-to-peer communication and collective communication.

First, all the peer-to-peer communication APIs, both synchronous and asynchronous operations, are implemented in the network model. The execution time of the MPI communication operations is independent of the message content. Hence, we model the performance based on the message size and the underlying network. Different communication protocols are used for different message sizes, such as ``eager'' or ``rendezvous''. Many state-of-the-art MPI simulators, such as SMPI~\cite{degomme_simulating_2017}, have depicted this design methodology and proven good simulation accuracy for a wide range of settings without any application-specific tuning. Our approach is similar, a linear model is used to predict the MPI communication performance. This model is built on top of the hardware model discussed in section~\ref{sec:hardware_model}. The network contention is simulated using the underlying network model. Figure~\ref{fig:mpi_send} illustrates the implementation of MPI send operation. At first, the global server IDs of the sender and receiver are obtained. Then, a network function $SendData$ is called to pass the communication request to the network model.
\begin{figure}
  \centering
\begin{lstlisting}[language=c++]
extern "C" int MPI_Send(const void *buf, int count, 
                        MPI_Datatype datatype,
                        int dest, int tag, 
                        MPI_Comm comm) {
    int src_id, dest_id;
    // Global ID of source and dest processes
    MPI_Comm_globalID(dest, comm, dest_id, src_id);
    double data_size = count * datatype;
    // This function returns after the data is sent
    SendData(src_id, dest_id, data_size);
    return MPI_SUCCESS;
}
\end{lstlisting}
  \caption{Implementation of MPI Send.}
  \label{fig:mpi_send}
\end{figure}

Second, we model collective communications. Previous studies show that the performance and scalability of MPI collective communication operations are critical to HPC applications. In major MPI implementations, each collective operation has several different algorithms to chose from depending on several factors, such as the message size and network topology. In some algorithms, collective communication is broken into a set of peer-to-peer operations. In our model, several algorithms for each operation are simulated mimicking the behavior of real implementations of OpenMPI and IntelMPI. In addition, optimized algorithms for specific network topologies, such as torus and dragonfly networks, are also available.

\subsection{Modeling applications behavior}

In a previous section, we discussed several approaches to model application behavior. One traditional approach is to study and analyze the application source code, mimic its behavior at an abstract level, and model its critical components. While this method offers a high modeling accuracy, it is time-consuming and requires frequent follow-up model refinements.

With the hardware infrastructure and libraries models, our goal is to enable the modeling of HPC applications with few modifications to the application source code instead of mimicking applications behavior.  To achieve this goal, several challenges need to be addressed. We use HPL as an example in this section.

\textbf{Parallel processes:} Our framework employs Intel CoFluent Studio (CoFluent)~\cite{intel_product_nodate} which provides an easy to use graphical modeling tool in a SystemC simulation environment. Since SystemC is a sequential simulation engine, every MPI process of the application needs to be mapped onto a SystemC thread.~\cite{bian_simulating_2014} describes how to mimic an application parallel behavior in detail. As the native application source code is used in our approach, each MPI process is bound with a SystemC virtual thread. Using this approach, all the HPL processes are simulated with low overheads.
%
%

\textbf{Integration of SimBLAS and SimMPI libraries:} The original HPL source code supports several BLAS interfaces, for example, CBLAS and FBLAS. Here, we enable SimBLAS interfaces in HPL source code. Only three modifications to the HPL source code are needed, defining SimBLAS and including the new header file. SimMPI supports the same APIs as the standard MPI library. Hence, enabling SimMPI in HPL source code is simply achieved by including a header file.

\textbf{Simulation of other components:} In addition to the BLAS computations and MPI communications, HPL spends significant time performing local copy and swap operations. In order to model HPL accurately, these HPL kernels, such as $HPL\_dlaswp*$, are simulated using the same approach used for BLAS Level-1 operations. Furthermore, $\min$ and $\max$ functions are simulated with random numbers as the content has no impact on HPL behavior. 

\textbf{Privatization of global variables:} As the CoFluent kernels are implemented in SystemC, which uses a single process to simulate parallel MPI processes, global variables in the application code are shared between all MPI processes. In our framework, a private copy of the global variables is stored for each parallel process. CoFluent offers a simple API, $get\_container()$, which can be used by a virtual thread to get the corresponding MPI rank. A global array is used to store the privatized variables and can be accessed using a dedicated index.

The last challenge is to identify which components of the source code to modify. In this work, optimizations for simulation speed are used to identify the modifications. The two largest data structures in HPL are matrix $A$ and the $panel$ which stores the workplace. The total space allocated by the MPI processes on each node typically consumes most of the node memory while the content of $A$ is irrelevant for the simulation. This memory allocation is removed with small modifications to code. The simulation results also indicate no impact on the execution flow and simulation accuracy.

Even though the matrix $A$ can be removed, $panel$ is used in every iteration of the factorization and, hence, must be stored. A possible workaround is to allocate and free $panel$ structure at every iteration. However, this option is time-consuming. Alternatively, we use a global array to store $panel$ structures for all MPI processes and $panel$ $init/free$ functions are reimplemented to map/demap corresponding spaces to private addresses.

\section{Performance Validation and Scalability Evaluation}
\label{sec:results}

In this section, we first discuss the accuracy of our framework. Then, we examine its scalability by performing simulations while changing the number of MPI processes from $2,000$ to $10,000$. Lastly, we demonstrate the fast simulation speed with different problem sizes and various configuration settings.

\begin{table}
  \caption{Hardware and Software configurations.}
  \label{tab:setup}
  \def\arraystretch{1.2}
  \begin{tabular}{c c}
\hline
    Category & Details\\
\hline
    \texttt{Node\#} & 4 \\
    \texttt{CPU\#}& Intel Xeon Broadwell E5-2699 v4 @ 2.2GHz \\
    \texttt{Socket\#}& 2\\
    \texttt{Cores\#/Socket}& 22\\
    \texttt{Memory/node}& DDR4 256GB @ 2400MHz\\
    \texttt{Network}& 1 Port Intel OPA 100Gb \\
\hline
    \texttt{HPL version}& Open HPL v2.3, Intel HPL v2.2\\
    \texttt{MPI version}& Intel MPI 2019\\
\hline
\end{tabular}
\end{table}

\subsection{Simulation accuracy}

To validate the simulation accuracy, we conduct experiments on our local environment. Table~\ref{tab:setup} shows the configurations details of the environment. The cluster has 4 nodes, each node has a dual-socket of Intel Xeon CPU with 22 cores per socket. Each node has 256GB DDR4 memory operating at frequency 2.4GHz. All nodes are connected to the same switch with a single port of Intel 100Gb OPA. Software configurations are also shown in Table~\ref{tab:setup}. Two HPL versions, OpenHPL 2.3 and Intel HPL 2.2 are installed. We choose the two versions since they are both widely used in supercomputing systems as demonstrated in the TOP500 list. OpenHPL is compiled with Intel MKL 2019 and Intel MPI 2019. Intel HPL is based on Open HPL 2.2 and is available as a part of the Intel MKL library. Both HPL implementations use the same hardware and same Intel MPI library.

OpenHPL uses one core per MPI process while Intel HPL uses all cores per node for each MPI process. Hence, the optimal $\mathrm{P} \times \mathrm{Q}$ combination for each HPL implementation is different, where P and Q are the rows and columns of the MPI process grid in the benchmark. This allows for more validation scenarios while having no impact on the validation process as we are not comparing the variance of the two HPL implementations. For the given architecture, the HPL block size used is $nb = 192$. The efficiency of the BLAS operations is evaluated using the methodology discussed in section~\ref{sec:blas}. The theoretical CPU peak performance and memory bandwidth are given as inputs to the simulator.

\begin{figure}
  \centering
  \includegraphics[width=\linewidth]{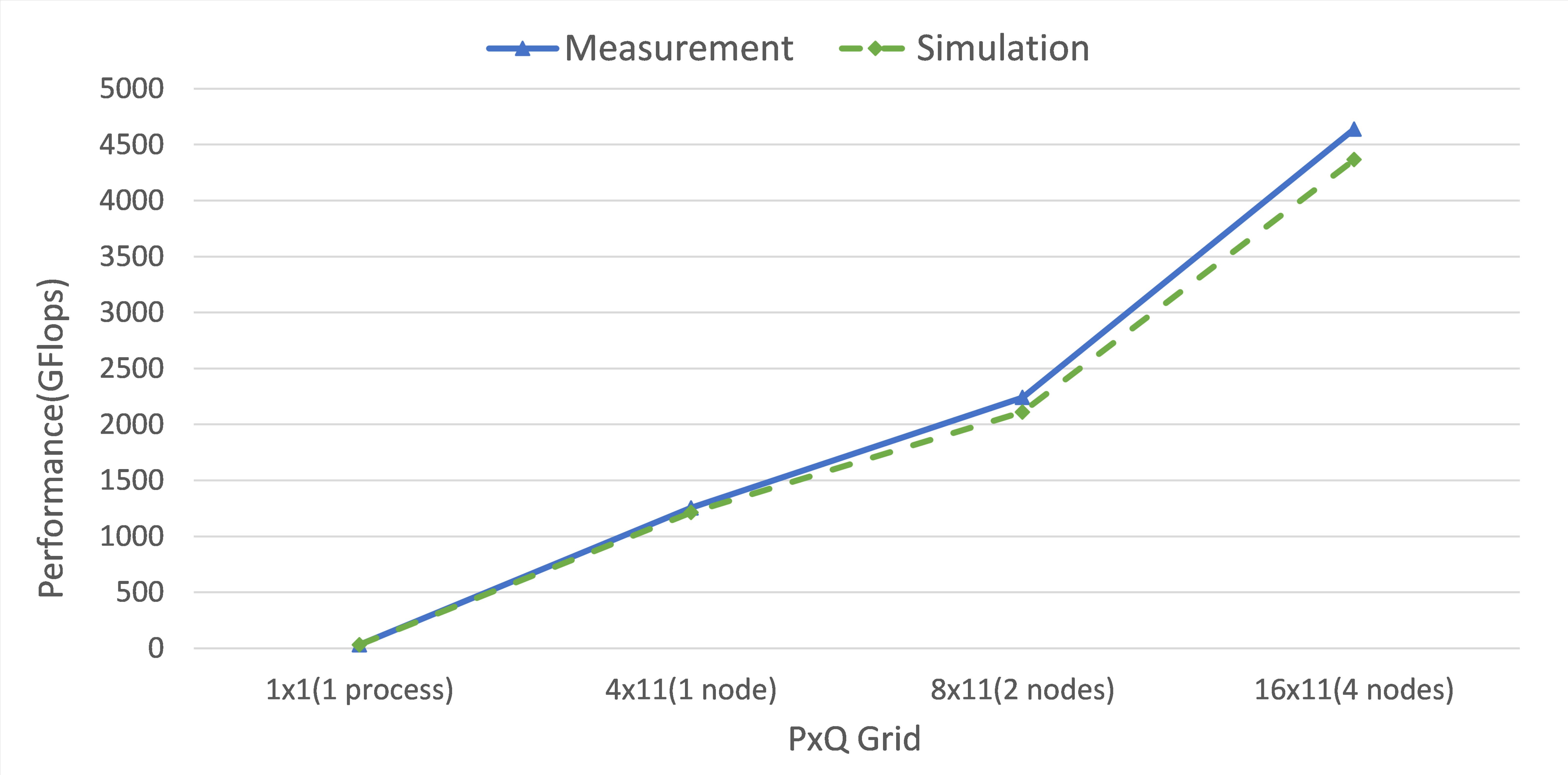}
  \caption{OpenHPL performance.}
  \label{fig:openhpl_accuracy}
\end{figure}
\begin{figure}
  \centering
    \includegraphics[width=\linewidth]{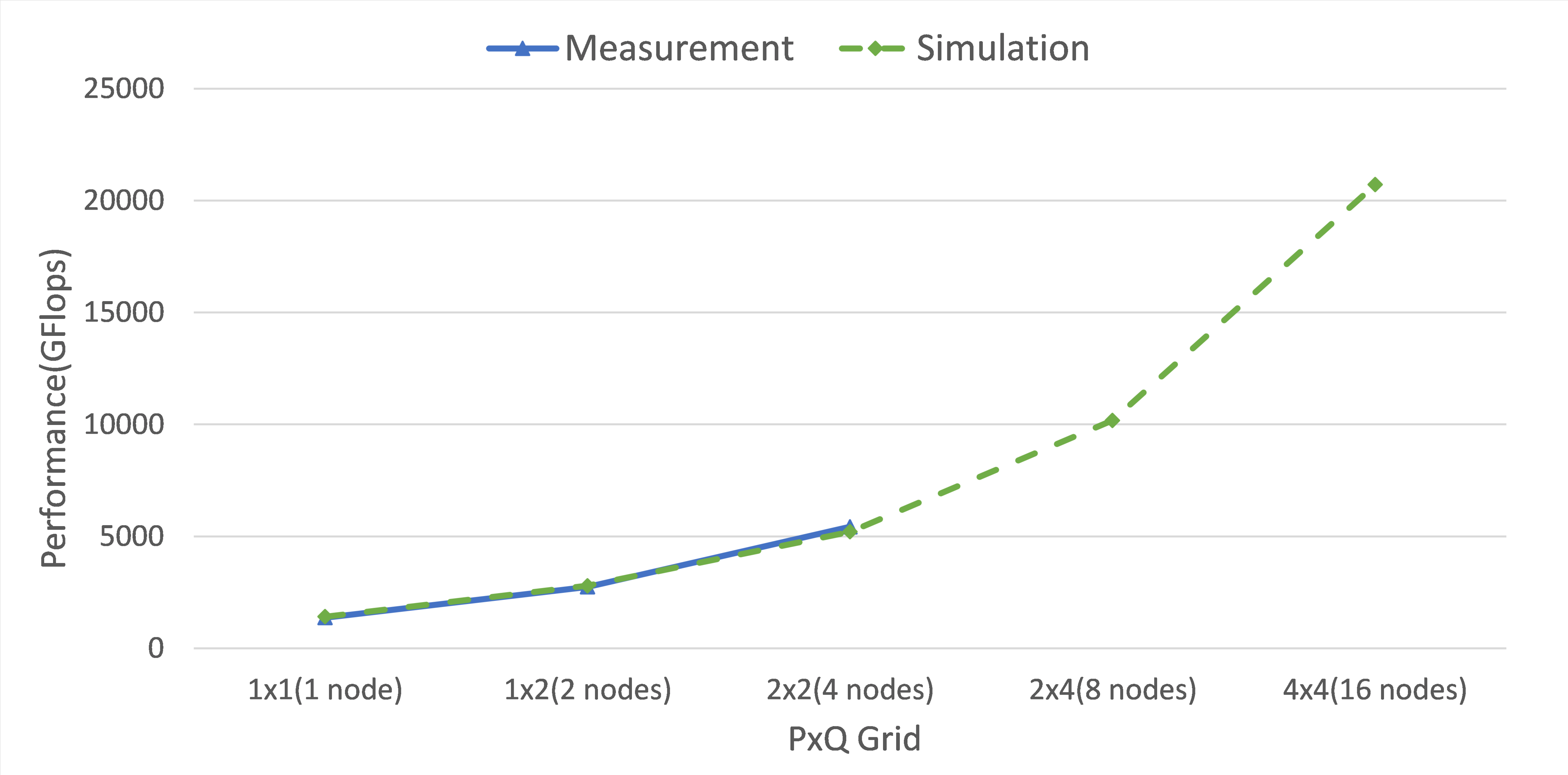}
  \caption{Intel HPL performance.}
  \label{fig:intelhpl_accuracy}
\end{figure}
Figure~\ref{fig:openhpl_accuracy} compares the simulated performance of OpenHPL against the measured performance on 1 core to 4 nodes. Figure~\ref{fig:intelhpl_accuracy} shows the validation results of Intel HPL with node numbers scaling from 1 to 4. The performance on 8 and 16 nodes is predicted using the simulator. Overall, our framework achieves high accuracy at varying concurrency with an average of $3.7\%$ discrepancy between the simulated and measured performance.

\subsection{Simulation scalability}

To evaluate the scalability of our framework, we simulate an HPC system consisting of 10,008 nodes. These nodes are connected using a two-level fat-tree topology. In total, 556 36-port switches are used at the edge level and 18 556-port switches are used at the core level. Each of the edge switches has 18 ports dedicated to connecting servers. The other 18 ports of each edge switch are connected to the core layer. In this scenario, the network of this hypothetical system may not be fully optimized as our goal is to evaluate the scalability of the simulator. The other hardware components are kept the same as those used for the experiments in the previous section.

\begin{figure}
  \centering
  \includegraphics[width=\linewidth]{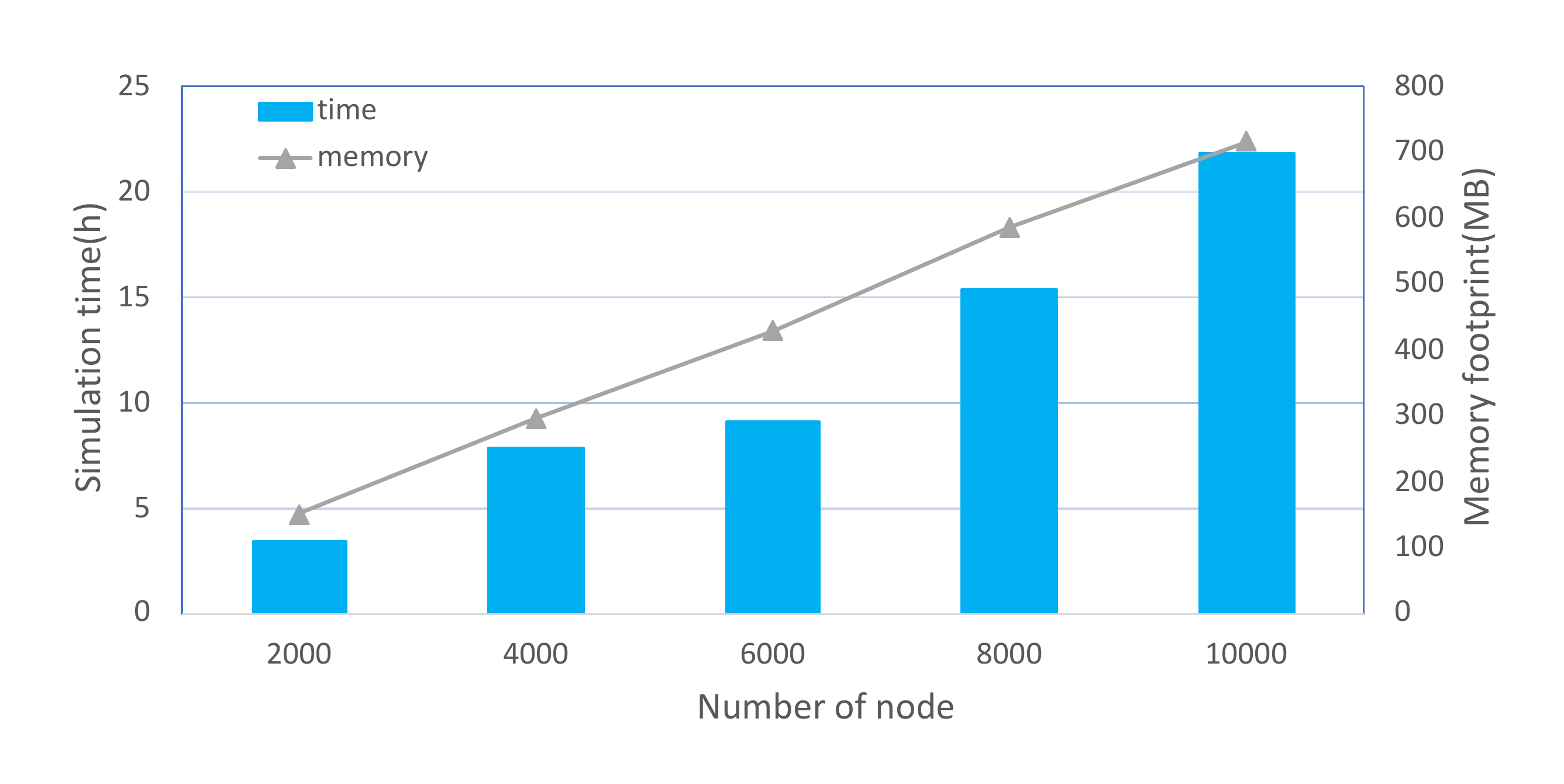}
  \caption{HPL simulation time and memory footprint.}
  \label{fig:scalability}
\end{figure}
The number of MPI processes and the matrix size are the two key factors impacting the HPL simulation time and memory consumption. In this section, we conduct a series of simulations where the matrix size is fixed to $2 \times 10^{7}$ while the number of MPI processes varies from $2,000$ to $10,000$ with a step size of $2,000$. The simulation results are shown in Figure~\ref{fig:scalability}. The bars in the figure represent the execution time. The largest simulation time is 21.8 hours which simulates $10,000$ MPI processes with a matrix size of $2 \times 10^{7}$. The line in Figure~\ref{fig:scalability} represents the memory footprint. The memory consumption grows linearly with the number of MPI processes. Simulating $10,000$ MPI processes consumes about 720MB.


\subsection{TOP500 HPC systems simulation}

The TOP500 list ranks the most powerful supercomputing systems according to their performance on the HPL benchmark. Frontera~\cite{noauthor_frontera_nodate} and PupMaya~\cite{noauthor_pupmaya_nodate} supercomputers, which rank \#5 and \#25 on the TOP500 list, respectively, provide enough public information to allow the use of our simulator to predict their HPL performance. 

Table~\ref{tab:TOP500} shows the hardware configurations along with the performance reported in the TOP500 list. Frontera consists of $8,008$ compute nodes, each node consists of a 2 socket Intel Xeon Platinum 8280 2.7GHz CPU with 28 cores per socket, and a 192GB DDR4 memory operating at frequency 2933 MHZ. One thing to note here is that the Cascade Lake processor cannot operate at 2.7GHz continuously when running 512-bit Advanced Vector Extensions (AVX-512) unit and the actual running frequency is around 1.8 GHz. The peak CPU performance, memory bandwidth, and kernels efficiency are given as inputs to the simulator. Furthermore, we configure the simulator to use Frontera's network topology which consists of six core switches, 182 leaf switches, and Mellanox HDR InfiniBand technology with 100Gbps and 90ns latency per routing hop~\cite{noauthor_system_nodate}, connected in a two-level fat-tree topology (Half of the nodes in a rack (44) connect to 22 downlinks of a leaf switch as pairs of HDR100 (100 Gb/s) links into HDR200 (200 Gb/s) ports of the leaf switch. The other 18 ports are uplinks to the six core switches). We assume that the routing algorithm is a non-blocking D-mod-K as it is commonly used in fat-tree networks~\cite{zahavi_d-mod-k_2010}. We also assume default MPI configurations.

\begin{table}
  \caption{TOP500 systems simulation.}
  \label{tab:TOP500}
  \def\arraystretch{1.2}
  \begin{tabular}{c c c c }
\hline
     & & Real environment & Simulation \\
\hline
    \multirow{6}{*}{\rotatebox{90}{Frontera}} & \texttt{Node\#} & 8,808 & 1 \\
    & \texttt{Core\#} & 448,448 & 1  \\
    & \texttt{Memory} & 1,537,536 GB & 550 MB \\
    & \texttt{Nmax}   & 9,282,848 & 9,282,848\\
    & \texttt{Rmax}   & 23,516 TFLOP/s & 22,566 TFLOP/s\\
    & \texttt{Execute time}& 6.5 h (Estimated) & 4.8 h\\
\hline
    \multirow{6}{*}{\rotatebox{90}{PupMaya}} & \texttt{Node\#} & 4,248 & 1 \\
    & \texttt{Core\#} & 169,920 & 1  \\
    & \texttt{Memory} & 815,616 GB & 300 MB \\
    & \texttt{Nmax}   & 4,748,928 & 4,748,928 \\
    & \texttt{Rmax}   & 7,484 TFLOP/s & 7,558 TFLOP/s\\
    & \texttt{Execute time}& 2.7 h (Estimated) & 1.7 h \\
\hline
\end{tabular}
\end{table}
The simulation results are shown in Table~\ref{tab:TOP500}. The simulated performance of Frontera is $22,566$ TFLOPs, while the Rmax performance reported in the TOP500 list is $23,516$ TFLOPs. The error rate is around $4\%$. The simulator execution time is 4.8 hours with about 550MB memory consumption, which is faster than the actual running time of more than 6.5 hours on the full-system (we estimate the actual time based on the problem size).

PupMaya consists of $4,248$ nodes, almost half the size of the Frontera supercomputer. We simulate the HPL performance on PupMaya using our framework and achieve good accuracy. Simulation results are shown in Table~\ref{tab:TOP500}.

\section{Use cases}
\label{sec:cases}

In this section, we use HPL as an example to demonstrate the simulation framework capabilities to perform what-if analysis.

In the previous section, the HPL performance on Frontera and PupMaya supercomputers is simulated. These two systems use Mellanox InfiniBand 100Gbps as their interconnect. Here, we use the simulator to predict the HPL performance on a 200Gbps network. Our simulation results show that the performance of Frontera increases from $22,566$ TFLOP to $23,143$ TFLOPs, and that of PupMaya increases from $7,558$ TFLOPs to $7,854$ TFLOPs. The performance improvement rates are $2.6\%$ and $3.9\%$ for Frontera and PupMaya, respectively, which are very low. A closer look at the simulation results shows that network congestion occurs due to the non-blocking routing algorithm used in the fat-tree network. Therefore, in this scenario, the high cost of updating the network does not lead to significant performance improvement.

A large portion of HPC systems on the TOP500 list are equipped with accelerators, such as GPGPU. It is therefore of interest to simulate heterogeneous systems to predict and optimize the performance of scientific applications on emerging large scale systems. HPL CUDA~\cite{avidday_aviddayhplcuda_2020} is an open-source HPL implementation for NVIDIA GPU. However, the code was last updated in 2011 and is based on HPL version 2.0. On our local server, this implementation achieved performance is about half the theoretical peak while both Summit~\cite{TOP500_summit_nodate} and Sierra~\cite{sierra_sierra_nodate} supercomputers achieve more than $75\%$ efficiency. Unfortunately, although we can correlate the simulator with local measurements, the low compute efficiency is far from practical use for predicting the performance of modern HPC systems.

\section{Conclusion}
\label{sec:conclusion}

The exponential increase in core counts expected at exascale will lead to increases in the number of switches,  interconnects,  and memory systems. For this reason, modeling application performance at these scales and understanding what changes need to be made to ensure continued scalability on future exascale architectures is necessary. 

This paper proposes a simulation approach to facilitate this process. Our approach enables full-system performance modeling: (1) the hardware platform is represented by an abstract yet high-fidelity model; (2) the computation and communication components are simulated at a functional level, where the simulator allows the use of the components native interface; this results in a (3) fast and accurate simulation of full HPC applications with minimal modifications to the application source code. This hardware/software hybrid modeling methodology allows for low overhead, fast, and accurate exascale simulation and can be easily carried out on a standard client platform (desktop or laptop). HPL is used to demonstrate the capability and scalability of the simulator. Two supercomputers from the TOP500, Frontera and PupMaya, are simulated with good simulation speed and accuracy. Specifically, the simulation of the HPL benchmark on Frontera takes less than 5 hours with an error rate of four percent.

We are extending our simulation framework in several ways to build a more comprehensive solution for modeling and exploiting the full performance of exascale computing platforms. Multithreading is widely used in HPC applications. In the current implementation, threads are extracted manually. We are working on automating this process in CoFluent Virtual Thread by enabling the simulation of Linux Pthreads and C++ threads. We also plan to support an automatic privatizing of the global variables when mapping applications processes into virtual threads. Finally, power is a major challenge for exascale systems. We are planning to incorporate power models into the simulation framework to enable the design of energy-efficient hardware and software.

\bibliographystyle{IEEEtran}
\bibliography{references}

\end{document}